\documentclass[12pt]{amsart}
\usepackage{amssymb}
\usepackage{amsmath}
\usepackage{bm}
\usepackage[dvips]{epsfig}
\usepackage{graphicx}
\usepackage{amsaddr}

\begin{document}

\hfill{Dedicated to the 85th birthday of Sergey Petrovich Novikov} 

\vspace{1cm}

\title{On the Novikov problem with a large number of quasiperiods 
and its generalizations}

\author{A.Ya. Maltsev}

\address{
\centerline{\it L.D. Landau Institute for Theoretical Physics 
of Russian Academy of Sciences}
\centerline{\it 142432 Chernogolovka, pr. Ak. Semenova 1A}
}

\begin{abstract}
The paper considers the Novikov problem of describing the geometry 
of level lines of quasi-periodic functions on the plane. We consider 
here the most general case, when the number of quasi-periods of a
function is not limited. The main subject of investigation is the 
arising of open level lines or closed level lines of arbitrarily 
large sizes, which play an important role in many dynamical systems 
related to the general Novikov problem. As can be shown also, 
the results obtained for quasiperiodic functions on the plane can be 
generalized to the multidimensional case. In this case, we are 
dealing with a generalized Novikov problem, namely, the problem of 
describing level surfaces of quasiperiodic functions in a space of 
arbitrary dimension. Like the Novikov problem on the plane, 
the generalized Novikov problem plays an important role in many 
systems containing quasiperiodic modulations.
\end{abstract}

\maketitle

\section{Introduction}
\setcounter{equation}{0}

 We consider here the Novikov problem, namely, the problem of 
describing the level lines of quasi-periodic functions on the plane 
with an arbitrary number of quasi-periods. In this case, the 
description of the global geometry of open (non-closed) level 
lines is of greatest interest, although the study of closed level 
lines also plays an important role in describing the general picture.
We will here call a quasi-periodic function on the plane 
$\, f (x^{1}, x^{2}) \, $ the restriction of an $N$-periodic 
function $\, F (z^{1}, \dots, z ^{N}) \, $ for some affine 
embedding $\, \mathbb{R}^{2} \subset \mathbb{R}^{N} \, $. The 
number $N$ is called the number of quasi-periods of the 
function $\, f (x^{1}, x^{2}) \, $. Here we will assume that 
the function $\, F (z^{1}, \dots, z^{N}) \, $ is sufficiently smooth.

 The first non-trivial and most deeply studied case of the 
Novikov problem is the description of the level lines of 
quasi-periodic functions on the plane with three quasi-periods. 
It is in this formulation that this problem was first set 
in \cite{MultValAnMorseTheory} and then intensively studied 
at Novikov's topological school. The Novikov problem with three 
quasiperiods plays a very important role in the theory of 
magnetoconductivity of normal metals, since in this formulation 
it is directly related to the description of the geometry of 
semiclassical electron trajectories on Fermi surfaces in the 
presence of external magnetic fields. The geometry of such 
trajectories, in turn, has the most significant influence on the 
behavior of magnetotransport phenomena in metals in the limit 
$\, B \rightarrow \infty \, $ (\cite{lifazkag,lifpes1,lifpes2,etm}).
The most significant results, which form the basis for the 
description of level lines of functions with three 
quasi-periods on the plane, were obtained in the papers 
\cite{zorich1,dynn1992,Tsarev,dynn1,DynnBuDA,dynn2,dynn3}. 
The study of the Novikov problem in this case makes it 
possible to determine important topological quantities, 
as well as to give a classification of all non-trivial regimes 
observed in transport phenomena in metals in strong magnetic 
fields (see, for example, 
\cite{PismaZhETF,UFN,BullBrazMathSoc,JournStatPhys}).

 At present, a number of deep topological results have also 
been obtained for the Novikov problem with four quasi-periods 
(see \cite{NovKvazFunc,DynNov}). It can be noted, however, that 
this problem has not been studied in as much detail as the Novikov 
problem with three quasi-periods. 

 In general, the Novikov problem with an arbitrary number of 
quasi-periods has many important applications, especially in the 
theory of complex two-dimensional systems. The study of the Novikov 
problem, especially the description of the geometry of non-compact 
level lines, is associated with very diverse aspects of the dynamics 
in such systems. Very often arising quasiperiodic functions play 
the role of potential energy in the problems under consideration. 
We will therefore call here quasiperiodic functions also quasiperiodic 
potentials. Most often, the occurrence of open level lines, 
or closed level lines of arbitrarily large sizes, has a significant 
impact on the behavior of transport phenomena that arise in the 
corresponding two-dimensional systems. It can be noted that the 
study of the Novikov problem with a large number of 
quasiperiods can also be considered as one of the approaches 
to the study of random potentials on the plane. It must be said 
that the study of the Novikov problem, in a wide variety of its 
aspects, currently represents a whole direction in topology and 
the theory of dynamical systems (see, for example, a review of 
recent results and literature in \cite{DynMalNovUMN}).

 In this paper we present a number of results related to 
the Novikov problem with an arbitrary number of quasiperiods.
More precisely, we prove here a number of statements about 
the occurrence of open (non-compact) level lines, as well as 
closed level lines of arbitrarily large diameter  for 
embeddings $\, \mathbb{R}^{2} \subset \mathbb{R}^{N } \, $ and 
functions $\, F (z^{1}, \dots, z^{N}) \, $ in general position. 
The assertions formulated here are, in fact, a further development 
of the results obtained in \cite{DynMalNovUMN} and related to the 
Novikov problem in its most general formulation. The results 
presented here refer mainly to the case $\, N \geq 4 \, $. 
The case $\, N = 3 \, $, as shown in 
\cite{zorich1,dynn1992,Tsarev,dynn1,DynnBuDA,dynn2,dynn3}, 
actually admits a more detailed description.

 In fact, as we will also show here, similar statements can be 
proven for the generalized Novikov problem, namely, the problem 
of describing the geometry of level surfaces of quasiperiodic 
functions in spaces of arbitrary dimension. Like the results 
for the Novikov problem on the plane, these statements can play 
an important role in many phenomena associated with quasiperiodic 
modulations in higher-dimensional systems.

\section{Sets of open and closed level lines of quasi-periodic 
functions on the plane}
\setcounter{equation}{0}

\begin{figure}[t]
\begin{center}
\includegraphics[width=\linewidth]{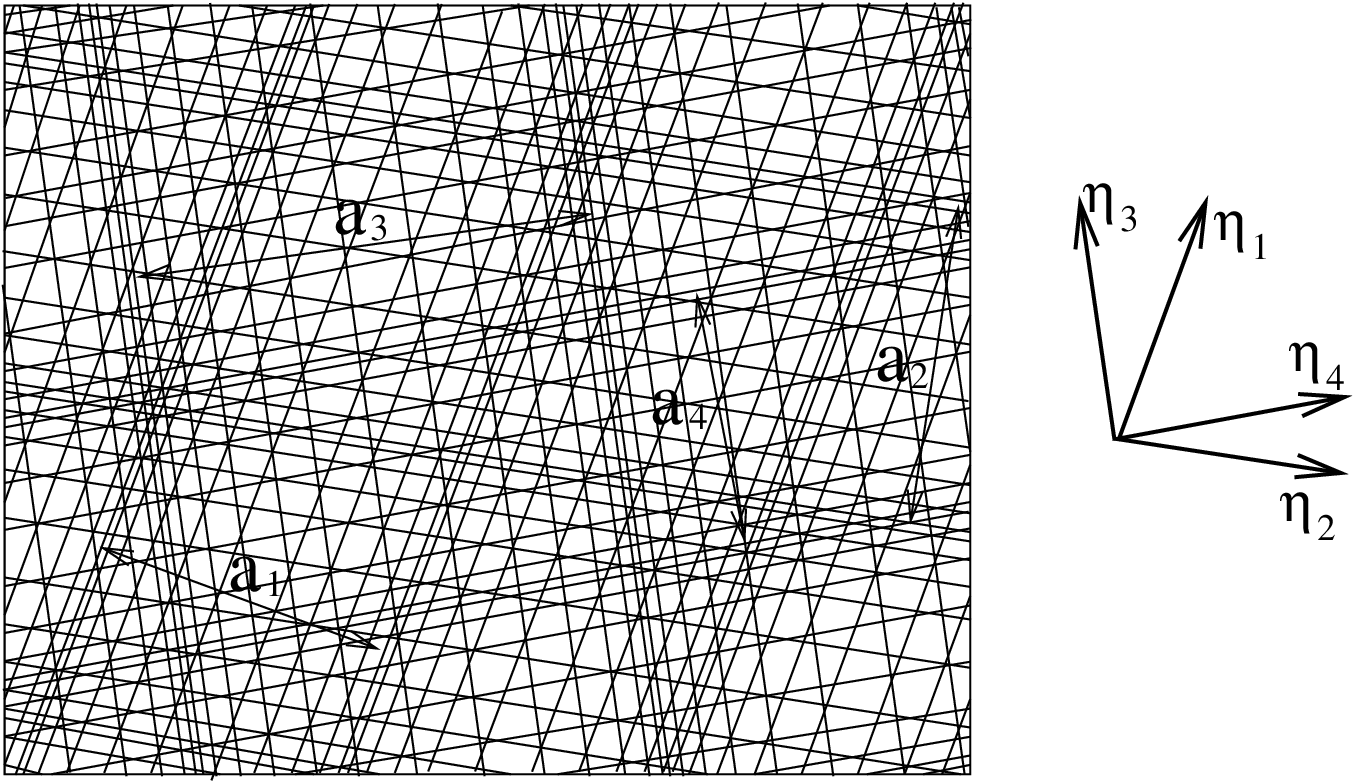}
\end{center}
\caption{Creation of quasiperiodic potentials on the plane 
by superposition of one-dimensional periodic potentials (schematically)
}
\label{Fig1}
\end{figure}

 In this work, we will consider the features of quasiperiodic 
potentials that appear in cases of four or more quasiperiods. 
In this formulation, Novikov’s problem does not, generally speaking, 
relate to the theory of galvanomagnetic phenomena in metals with 
complex Fermi surfaces. Instead, however, it is interesting in many 
different two-dimensional systems, where quasiperiodic potentials 
are created using certain experimental techniques. The most common 
technique in this case is, apparently, the method of superposition 
of several one-dimensional (or, more generally, periodic) potentials 
(Fig. \ref{Fig1}), the periods of which are incommensurate and 
differently oriented in the plane (see, for example, 
\cite{Guidoni2,SanchezPalenciaSantos,Guidoni3}). As is easy 
to show, the resulting potentials can also be obtained as the 
restriction of an $N$-periodic function to the plane for some 
embedding $\, \mathbb{R}^{2} \subset \mathbb{R}^{N} \, $ 
for a suitable value of $N$. In addition, it can also be noted 
that quasiperiodic (quasicrystalline) structures, for example, 
in systems of ultracold atoms, can arise also in the absence of 
special external modulation (see, for example, \cite{GopMartinDemler}). 
As in the case of three quasiperiods, Novikov’s problem here relates, 
as a rule, primarily to transport phenomena in such systems.

 Here it should immediately be noted that for functions with 
a large number of quasiperiods, even very general questions, 
such as questions of the presence of open level lines or closed 
level lines of arbitrarily large sizes, become nontrivial. 
At the same time, it is questions of this type that are primarily 
related to transport or, for example, percolation phenomena in 
the corresponding systems. Here we will consider some questions 
of this type from the most general point of view, using general 
constructions of the theory of quasiperiodic functions. We will 
now move on to such consideration.

\vspace{1mm}

 Consider an arbitrary affine embedding 
$\, \mathbb{R}^{2} \rightarrow \mathbb{R}^{N} \, $. Since 
for our purposes the affine transformations of the space 
$\, \mathbb{R}^{2} \, $ itself will not be of interest, the 
essential parameters for us here, in fact, are the 
two-dimensional direction $\, \xi \, $, and also the shift 
parameters of the space $\, \mathbb{R}^{2} \, $ in directions 
orthogonal to it in the space $\, \mathbb{R}^{N} \, $. 
The first part of these parameters is then the 
Grassmann manifold $\, G_{N,2} \, $, while the second one 
can be considered as the Euclidean space of dimension 
$\, N - 2 \, $. As in \cite{DynMalNovUMN}, it will be 
convenient here to consider at once the entire collection 
of parallel subspaces $\, \mathbb{R}^{2} \, $ in 
$\, \mathbb{R}^{N} \, $ for a fixed direction $\, \xi \, $.

 We will everywhere assume here that the direction $\, \xi \, $ 
is not rational and, moreover, is not contained in any integral 
hyperplane in $\, \mathbb{R}^{N} \, $. In addition, we will divide 
all such directions $\, \xi \, $ into partially irrational and 
completely irrational. Namely, we will call a direction $\, \xi \, $ 
partially irrational if it contains exactly one (up to a factor) 
integer vector from $\, \mathbb{R}^{N} \, $. We will call a direction 
$\, \xi \, $ completely irrational if it does not contain integer 
vectors from $\, \mathbb{R}^{N} \, $. For simplicity, we will here 
call non-special directions $\, \xi \, $ completely irrational 
directions that do not lie in any integer hyperplane 
in $\, \mathbb{R}^{N} \, $.

 It is easy to see that the description of the level lines 
$\, F |_{\Pi} = c \, $ in all planes $\, \Pi \, $ of 
direction $\, \xi \, $ is equivalent to the description of all 
level lines of the foliation defined on the surface $\, F ({\bf z}) = c \, $ 
in $\, \mathbb{R}^{N} \, $ by the direction $\, \xi \, $.

 It must be said that families of functions $\, F |_{\Pi} \, $ 
that arise in different planes of a given direction $\, \xi \, $ 
often also correspond to ``related'' potentials that arise in 
physical systems. In particular, in two-dimensional electronic 
systems or systems of cold atoms, where such potentials arise 
as a result of the superposition of periodic 
(or even one-dimensional) potentials, such families represent 
potentials arising at the same parameters and orientation of 
the initial potentials, except for their shifts in the 
corresponding plane. Due to the peculiarities of this technique, 
various potentials of the described family often arise during 
the experiment as a result of a shift in the maxima and minima 
of the initial potentials with other parameters fixed. For 
completely irrational directions $\, \xi \, $, it is often 
assumed that the properties of such potentials are not 
fundamentally different and there is no need to make much 
difference between them. In fact, as it turns out, even the 
fact of the presence of open level lines at important ``critical'' 
levels in such potentials may depend on the described shifts of 
the initial potentials, which corresponds to different pictures 
in different planes $\, \Pi \subset \mathbb{R} ^{N} \, $ of 
the same direction. Here we will discuss, in particular, 
these questions in the most general formulation.

 Here we consider the function $\, F ({\bf z}) \, $ to be sufficiently 
smooth generic $\, N$ - periodic function taking values in the 
interval $\, [F_{\min} , F_{\max } ] \, $.

 In the case $\, N = 3 \, $ (see, for example, \cite{dynn3} and 
references therein), the level line patterns arising in different 
planes of the same non-special direction $\, \xi \, $ are 
qualitatively the same. In particular, if open level lines 
appear at some level $\, c\, $ in one of these planes, they also 
appear in all other planes of this direction. In the case 
$\, N \geq 4 \, $, however, the situation at the same level 
$\, c \, $ may be different in different planes of a fixed 
(non-special) direction. Let us present here a number of results 
formulated in \cite{DynMalNovUMN} for the most general case.

\vspace{1mm}

\noindent
1) For any two-dimensional direction $\, \xi \in G_{N,2} \, $ 
the levels $\, c \, $ corresponding to the presence of unbounded 
connected components of the set $\, F |_{\Pi} = c \, $ in at 
least one plane $\, \Pi \, $ of direction $\, \xi \, $ form a 
connected closed interval $\, [c_{1} (\xi) , c_{2} (\xi) ]\, $,
$$F_{\min} \, < \, c_{1} (\xi) \, \leq \, c_{2} (\xi) \, < \, F_{\max} \,\,\, , $$
which can contract to a single point 
$\, c_{0} (\xi) = c_{1} (\xi) = c_{2} (\xi) \, $.

\vspace{1mm}

\noindent
2) For any non-special direction $\, \xi \, $ and any 
$\, c \in [c_{1} (\xi) , c_{2} (\xi)] \, $ each plane $\, \Pi \, $ 
of direction $\, \xi \, $ contains either unbounded connected 
components of the set $\, F |_{\Pi} = c \, $, or bounded connected 
components of this set of arbitrarily large size (or both).

\vspace{1mm}

\noindent
3) For any $\, c \notin [c_{1} (\xi) , c_{2} (\xi)] \, $ the sizes 
of all connected components of the set $\, F |_{\Pi} = c \, $ in all 
planes $\, \Pi \, $ of direction $\, \xi \, $ are limited by one 
constant $\, D (\xi , c) \, $ (which can grow indefinitely as 
$\, c \, $ approaches the interval $\, [c_{1} (\xi) , c_{2} (\xi)]$).

\vspace{1mm}

 Note that for $\, N = 3 \, $ assertion (3) is also true for bounded 
connected components of the set $\, F |_{\Pi} = c \, $ for all values 
of $\, c \in [F_{\min} , F_{\max}] \, $. In the general case, however, 
this is not the case.

\vspace{1mm}

\begin{figure}[t]
\begin{center}
\includegraphics[width=\linewidth]{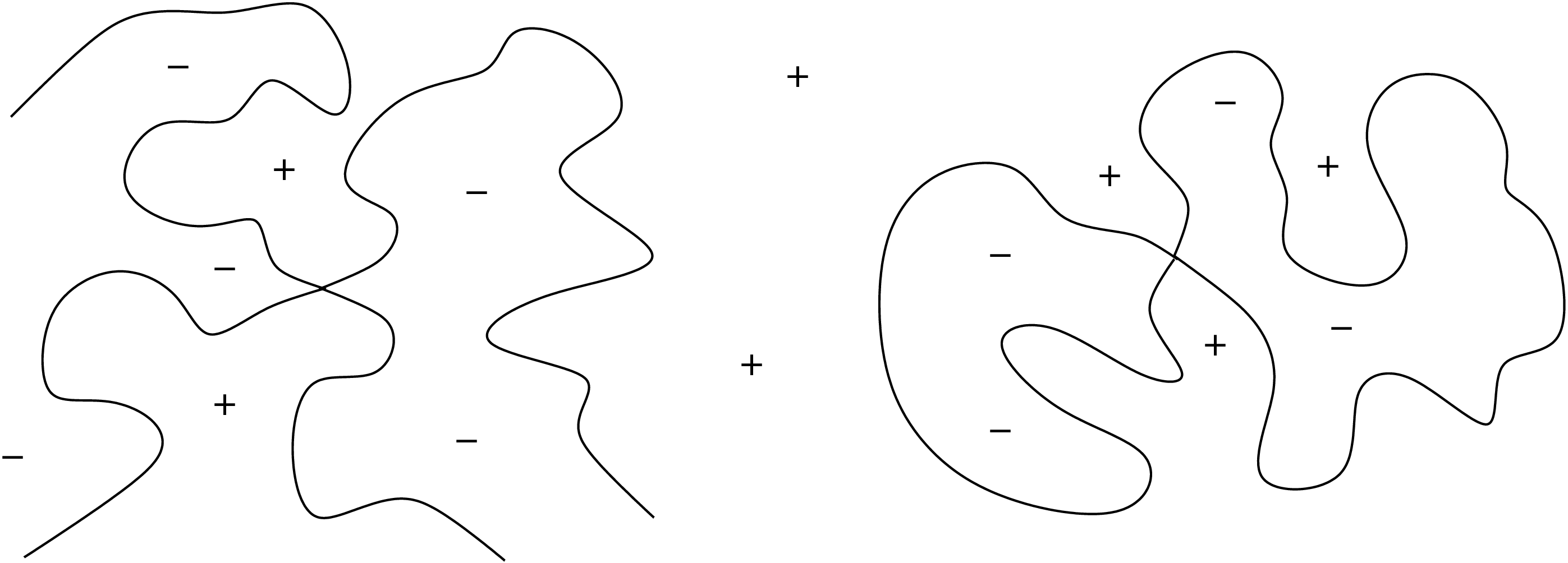}
\end{center}
\caption{Non-compact and compact singular connected components 
of the level $\, F |_{\Pi} = c \, $ (schematically)
}
\label{Fig2}
\end{figure}

 Here we prove a number of statements that complement the general 
picture considered in \cite{DynMalNovUMN}. Note here that the 
assertions in \cite{DynMalNovUMN} applied not only to non-singular but 
also to singular level lines $\, F |_{\Pi} = c \, $ 
(see Fig. \ref{Fig2}). It was assumed in this case that the 
functions $\, F |_{\Pi} \, $ can only have singularities 
of a certain form (multiple saddles, as well as isolated local maxima 
and minima). Moreover, the general position conditions 
in \cite{DynMalNovUMN} also assumed that for completely irrational 
directions $\, \xi \, $ any connected component of the 
set $\, F |_{\Pi} = c \, $ contains at most finitely 
many singular points of the function $\, F |_{\Pi} \, $. Here we also 
assume that these generic conditions hold.

 For simplicity, we will consider here the picture that arises in the 
planes $\, \Pi \, $, in which the levels $\, F |_{\Pi} = c \, $ are 
non-singular. Here we will call such planes non-singular planes. It is 
easy to see that for a generic function $\, F \, $ and any value 
of $\, c \, $ the non-singular planes of direction $\, \xi \, $ 
form a set of the full measure in any bounded domain in the set of 
planes of direction $\, \xi \, $ (that is, in the 
space $\, \mathbb{R}^{N-2}$).

\begin{figure}[t]
\begin{center}
\includegraphics[width=\linewidth]{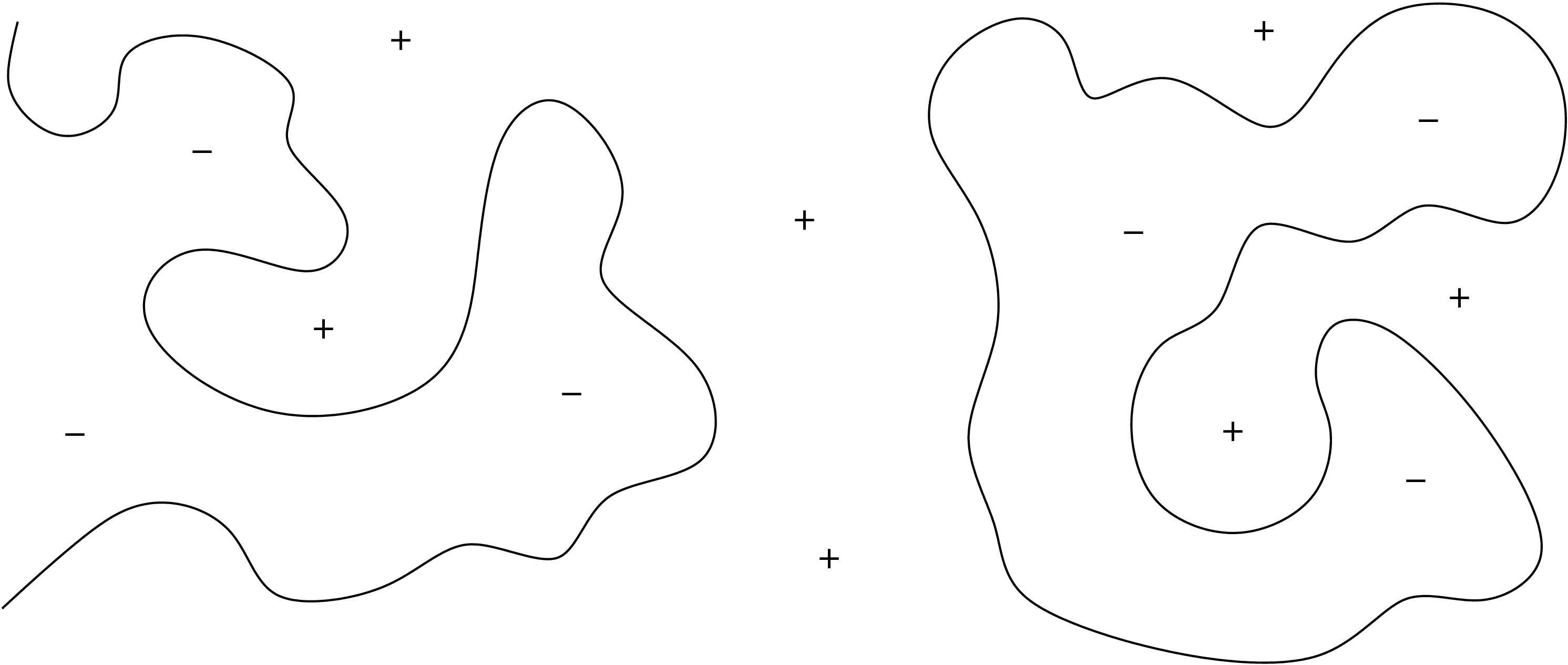}
\end{center}
\caption{Open and closed regular level lines 
$\, F |_{\Pi} = c \, $ (schematically)}
\label{Fig3}
\end{figure}

 The level lines are here the connected components of the set 
$\, F |_{\Pi} = c \, $ in non-singular planes for the value $\, c\, $. 
All level lines, therefore, will represent here non-singular curves, 
which we will divide into open (non-closed) and closed ones
(see Fig. \ref{Fig3}).

\vspace{1mm}

 When considering dynamics in physical systems, an important 
role can be played by both the geometry of potential level 
lines $\, F |_{\Pi} \, $ and the geometry of regions of higher 
or lower energies for a given energy level. We can see that 
in reality the geometric properties of these objects are 
interconnected. Here we will consider both the level lines 
of quasiperiodic potentials and the areas of their higher 
or lower values, as well as the connection of these objects 
with each other.

\vspace{1mm}

 For each plane $\, \Pi \, $ of direction $\, \xi \, $ and a fixed 
value $\, c \, $ denote by $\, \Omega^{-}_{c} (\Pi) \, $ the set 
of points in $\, \Pi \, $, where $\, F|_{\Pi} < c \, $. Similarly, 
denote by $\, \Omega^{+}_{c} (\Pi) \, $ the set of points in $\, \Pi \, $, 
where $\, F|_{\Pi} > c \, $.

 For each plane $\, \Pi \, $ we also introduce the notation 
$\, \sigma_{\bf a} [\Pi] \, $ for the parallel shift of $\, \Pi \, $ 
in the space $\, \mathbb{ R}^{N} \, $ by a vector 
$\, {\bf a} \in \mathbb{R}^{N} \, $. It is easy to see that integer 
shifts of any plane $\, \Pi \, $ of an irrational direction 
$\, \xi \, $ form in this case an everywhere dense set among the 
planes of the given direction. Similarly, for any set 
$\, M \subset \mathbb{R}^{N} \, $, we denote by 
$\, \sigma_{\bf a} [M] \, $ its parallel shift by the 
vector $\, {\bf a} \in \mathbb{R}^{N} \, $.

 According to \cite{DynMalNovUMN} (see detailed proof 
in arXiv:2306.11257) for any $\, c \in [c_{1} (\xi) , c_{2} (\xi) ] \, $ 
any plane $\, \Pi \, $ of a non-special direction $\, \xi \, $ contains 
either unbounded connected components of the set $\, F |_{\Pi} = c \, $, 
or bounded connected components of this set of arbitrarily large sizes. 
This is true, in particular, for the planes of direction $\, \xi \, $, 
that are non-singular for the value $\, c \, $. In this case, in such 
plane $\, \Pi \, $ there are always either open lines of the level 
$\, F |_{\Pi} = c \, $ or closed lines of this level of arbitrarily 
large sizes (or both).

 For non-singular planes $\, \Pi \, $ it is not difficult to prove the 
following Lemma.

\vspace{1mm}

\noindent
{\bf Lemma 2.1}

 Let $\, c \in [c_{1} (\xi) , c_{2} (\xi) ] \, $ and $\, \Pi \, $ 
be a plane of the direction $\, \xi \, $, non-singular for the 
value $\, c \, $. Then the presence of open level lines 
$\, F |_{\Pi} = c \, $ is equivalent to the existence of unbounded 
connected components of both sets $\, \Omega^{-}_{c} (\Pi) \, $ 
and $ \, \Omega^{+}_{c} (\Pi) \, $.

\vspace{1mm}

 Proof.

 Indeed, in the presence of an open level line $\, F |_{\Pi} = c \, $ 
we have two unbounded connected components of the sets 
$\, \Omega^{-}_{c} (\Pi) \, $ and $\, \Omega^{+}_{c} (\Pi) \, $ 
adjacent to it from different sides.

 Conversely (see e.g. \cite{dynn3}), let a plane $\, \Pi \, $ have 
unbounded connected components $\, M^{-} \, $ and $\, M^{+} \, $ of the 
sets $\, \Omega^{-} _{c} (\Pi) \, $ and $\, \Omega^{+}_{c} (\Pi) \, $ 
respectively. Let $\, {\bf x} \in M^{-} \, $ and 
$\, {\bf y} \in M^{+} \, $. Then $\, {\bf x} \, $ and $\, {\bf y} \, $ 
do not lie inside any of the closed level lines $\, F |_{\Pi} = c \, $.
Let us join $\, {\bf x} \, $ and $\, {\bf y} \, $ by a generic curve 
$\, \gamma \, $ in the plane $\, \Pi \, $. The curve 
$\, \gamma \, $ can only intersect a finite number of closed level 
lines in $\, \Pi \, $, while its intersection index with each of these 
lines is equal to zero. Since $\, \gamma \, $ must cross the level 
$\, F |_{\Pi} = c \, $ an odd number of times, it must intersect an 
open level line $\, F |_{\Pi} = c \, $.

\hfill{Lemma 2.1 is proved.}

\vspace{1mm}

 Let us call here a closed level line $\, F |_{\Pi} = c \, $ 
an ``electronic'' level line if the region 
$\, \Omega^{+}_{c} (\Pi ) \, $ adjoins it from the outside. 
Similarly, we call a closed level line 
$\, F |_{\Pi} = c \, $ a ``hole'' level line if the region 
$\, \Omega^{-}_{c} (\Pi)  \, $ adjoins it from the outside.

\vspace{1mm}

\noindent
{\bf Lemma 2.2}

 Let $\, c \in [c_{1} (\xi) , c_{2} (\xi) ] \, $, $\, \Pi \, $ 
is a plane of the corresponding non-special direction $\, \xi \, $, 
non-singular for the value $\, c \, $, and the size of the 
electronic level lines $\, F |_{\Pi} = c \, $ in the plane 
$\, \Pi \, $ is bounded above by one constant. Then 
in $\, \Pi \, $ there are unbounded connected components of 
the set $\, \Omega^{-}_{c} (\Pi) \, $.

\vspace{1mm}

 Proof.

 Indeed, if $\, c \in [c_{1} (\xi) , c_{2} (\xi) ] \, $, 
then $\, \Pi \, $ contains either open level lines 
$\, F |_{\Pi} = c \, $ (then everything is proved), or hole 
level lines $\, F |_{\Pi} = c \, $ of arbitrarily large sizes. 
In the latter case, $\, \Pi \, $ have connected components of 
the set $\, \Omega^{-}_{c} (\Pi) \, $ of arbitrarily large 
sizes (adjacent to large hole level lines). Since the size 
of the bounded connected components of the set 
$\, \Omega^{-}_{c} (\Pi) \, $ is bounded by one constant, 
$\, \Pi \, $ must contain an unbounded (connected) component 
of the set $\, \Omega^ {-}_{c} (\Pi) \, $.

\hfill{Lemma 2.2 is proved.}

\vspace{1mm}

 Similarly, there is

\vspace{1mm}

\noindent
{\bf Lemma 2.2$^{\prime}$}

  Let $\, c \in [c_{1} (\xi) , c_{2} (\xi) ] \, $, $\, \Pi \, $ 
is a plane of the corresponding non-special direction $\, \xi \, $, 
non-singular for the value $\, c \, $, and the size of the 
hole level lines $\, F |_{\Pi} = c \, $ in the plane 
$\, \Pi \, $ is bounded above by one constant. Then 
in $\, \Pi \, $ there are unbounded connected components of 
the set $\, \Omega^{+}_{c} (\Pi) \, $.

\vspace{1mm}

 On the whole, we can conclude that for a plane $\, \Pi \, $ 
of a non-special direction $\, \xi \, $, which is non-singular for 
a value $\, c \in [c_{1 } (\xi) , c_{2} (\xi) ] \, $ 
only one of the following four situations is possible.

\vspace{1mm}

\noindent
(A) The plane $\, \Pi \, $ contains open level lines 
$\, F |_{\Pi} = c \, $, as well as unbounded connected components 
of both sets $\, \Omega^{-}_ {c} (\Pi) \, $ 
and $\, \Omega^{+}_{c} (\Pi) \, $.

\vspace{1mm}

\noindent
B) In the plane $\, \Pi \, $, only the set 
$\, \Omega^{+}_{c} (\Pi) \, $ has unbounded connected components 
and there are no open level lines $\, F |_ {\Pi} = c \, $. In 
this case $\, \Pi \, $ must contain arbitrarily large closed level 
lines $\, F |_{\Pi} = c \, $ of the electronic type.

\vspace{1mm}

\noindent
(C) In the plane $\, \Pi \, $, only the set 
$\, \Omega^{-}_{c} (\Pi) \, $ has unbounded connected components 
and there are no open level lines $\, F |_ {\Pi} = c \, $. In 
this case $\, \Pi \, $ must contain arbitrarily large closed level 
lines $\, F |_{\Pi} = c \, $ of the hole type.

\vspace{1mm}

\noindent
(D) In the plane $\, \Pi \, $, there are no unbounded connected 
components of both sets $\, \Omega^{-}_{c} (\Pi) \, $ and 
$\, \Omega^{+}_ {c} (\Pi) \, $, as well as open level lines 
$\, F |_{\Pi} = c \, $. In this case $\, \Pi \, $ must contain 
arbitrarily large closed level lines $\, F |_{\Pi} = c \, $ of 
both electronic and hole types.

\vspace{1mm}

\begin{figure}[t]
\begin{center}
\includegraphics[width=\linewidth]{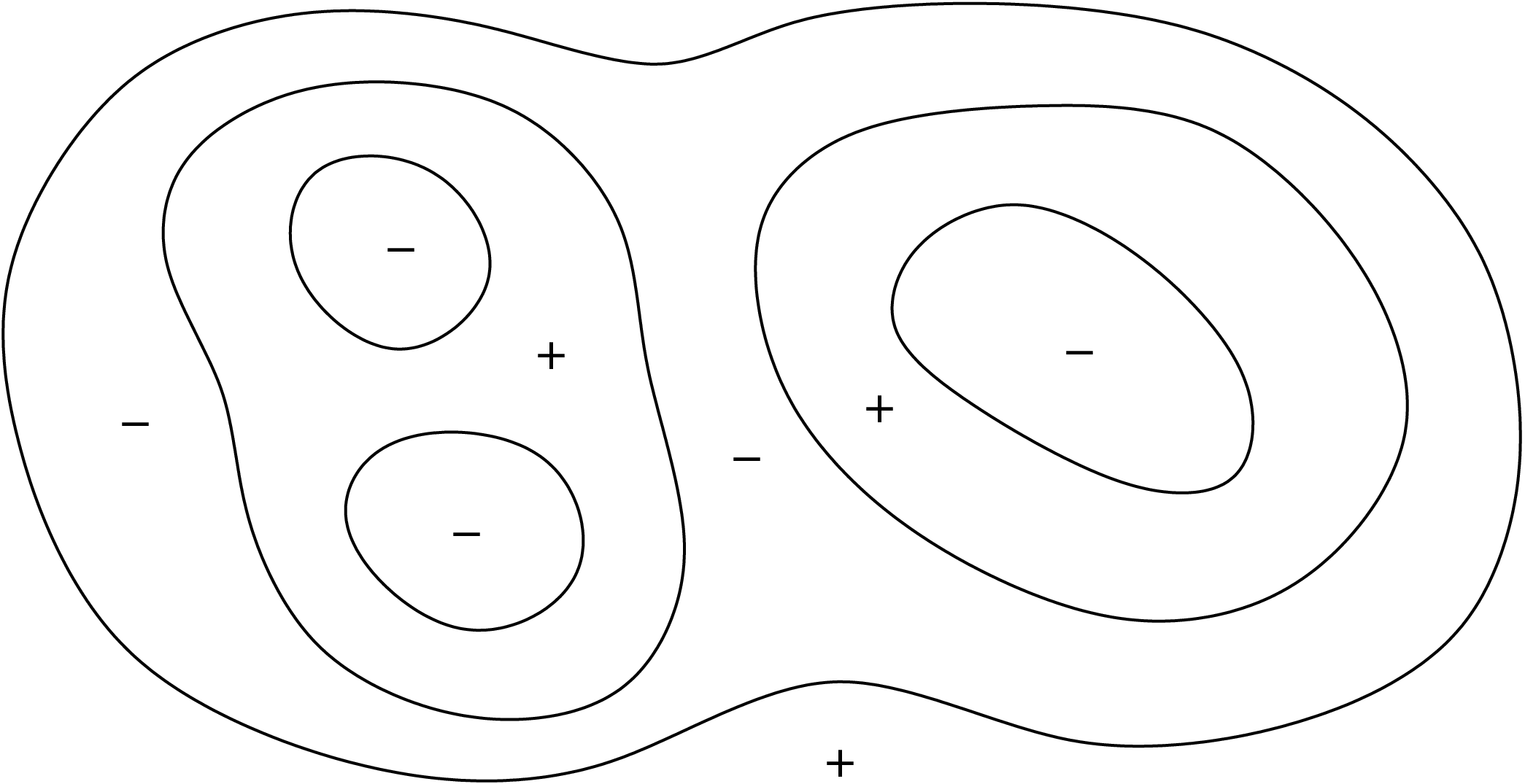}
\end{center}
\caption{Situation (D) in the plane $\, \Pi \, $ (schematically)}
\label{Fig4}
\end{figure}

 It is also easy to see that in case (D), due to the absence of 
unbounded connected components of the sets 
$\, \Omega^{-}_{c} (\Pi) \, $ and $\, \Omega^{+}_{ c} (\Pi) \, $, 
each electronic level line $\, F |_{\Pi} = c \, $ must be contained 
in some larger hole level line. In turn, each hole level 
line $\, F |_{\Pi} = c \, $ must also be contained in some 
larger electronic level line $\, F |_{\Pi} = c \, $ (etc.) 
(Fig. \ref{Fig4}).

\vspace{1mm}

\vspace{1mm}

\noindent
{\bf Theorem 2.1}

 Let $\, \xi \, $ be a non-special two-dimensional direction 
in $\, \mathbb{R}^{N} \, $, 
$\,\, c \in [c_{1} (\xi) , c_{ 2} (\xi) ] \, $, 
and $\, \Pi \, $ be a plane of direction $\, \xi \, $, non-singular 
for the value $\, c \, $. Then in the plane $\, \Pi \, $:

\vspace{1mm}

\noindent
1) Situation (B) is possible only for $\, c = c_{1} (\xi) \, $.

\vspace{1mm}

\noindent
2) Situation (C) is possible only for $\, c = c_{2} (\xi) \, $.

\vspace{1mm}

\noindent
3) Situation (D) is possible only for 
$\, c = c_{0} (\xi) = c_{1} (\xi) = c_{2}(\xi) \, $.

\vspace{1mm}

 Proof.

 Since the function $\, F ({\bf z}) \, $ can also be considered 
as a smooth function on the torus $\, \mathbb{T}^{N} \, $, there 
is a constant $\, C \, $, such that 
$\, |\nabla F ({\bf z})| \leq C \, $ 
for all $\, {\bf z} \in \mathbb{R}^{N} \, $.

 Let in some plane $\, \Pi \, $ of direction $\, \xi \, $, 
non-singular for the value $\, c \in [c_{1} (\xi) , c_{2} (\xi) ] \, $, 
there are no unbounded connected components of the 
set $\, \Omega^{-}_{c} (\Pi) \, $. In this case, each connected 
component of the set $\, \Omega^{-}_{c} (\Pi) \, $ is bounded by 
some closed level line $\, F |_{\Pi} = c \, $ in the plane $\, \Pi\, $.
Consider an arbitrary level $\, c^{\prime} < c \, $ and all 
parallel shifts $\, \sigma_{\bf a} [\Pi] \, $ of the plane 
$\, \Pi \, $ in $\, \mathbb{R}^{N} \, $ by vectors $\, {\bf a} \, $.

 It is easy to see that for $\, |{\bf a}| < (c - c^{\prime})/C \, $ 
the set $\, \Omega^{-}_{c^{\prime}} (\sigma_{\bf a} [\Pi]) \, $ belongs 
to the set $\, \sigma_{\bf a} [ \Omega^{-}_{c} (\Pi) ] \, $, and thus 
each of its connected components lies inside some non-singular closed 
curve in the plane $ \, \sigma_{\bf a} [\Pi] \, $. As a consequence, 
each connected component of the boundary of this set, i.e. the 
set $\, F|_{\sigma_{\bf a}[\Pi]} = c^{\prime} \, $, is also bounded 
by some non-singular closed curve that does not intersect other such 
curves in the plane $\, \sigma_{\bf a} [\Pi] \, $.

 Thus, in all planes $\, \sigma_{\bf a} [\Pi] \, $ for 
$\, |{\bf a}| < (c - c^{\prime})/C \, $ there are no unbounded 
connected components (nonsingular or singular) of the 
set $\, F|_{\sigma_{\bf a}[\Pi]} = c^{\prime} \, $. 
Since $\, \xi \, $ is a non-special direction 
in $\, \mathbb{R}^{N} \, $, any plane of direction $\, \xi \, $ 
coincides with one of these planes $\, \sigma_{\bf a} [\Pi] \, $ 
after a parallel shift by some integer vector from 
$\, \mathbb{R}^{N} \, $. We thus obtain 
that $\, c^{\prime} \notin [c_{1} (\xi) , c_{2} (\xi) ] \, $.

 Similarly, one can prove that if in some plane $\, \Pi \, $ 
of a non-special direction $\, \xi \, $, non-singular for a 
value $\, c \in [c_{1} (\xi) , c_{ 2} (\xi) ] \, $, there 
are no unbounded connected components of the set 
$\, \Omega^{+}_{c} (\Pi) \, $, then any value 
$\, c^{\prime} > c \, $ does not belong to the 
interval $\, [c_{1} (\xi) , c_{2} (\xi) ] \, $. 
All statements of the theorem follow directly from these
statements.

\hfill{Theorem 2.1 is proved.}

\vspace{1mm}

 One substantial remark can be made here. Namely, 
quasiperiodic potentials with a large number of 
quasiperiods can be considered, among other things, 
as a model of random potentials on the plane 
(possessing, however, some long-range order). 
It can be noted here that the use of quasiperiodic 
potentials for modeling random potentials plays 
an important role not only in the multidimensional, 
but also in the one-dimensional case (see, for example, 
\cite{WZLWLMHXMCJ,LesserLifshitz} and the references given 
there). Here we would also like to discuss aspects of this 
approach in the case of two dimensions.

 As a rule (see, for example, \cite{Stauffer,Essam}), 
in the theory of random potentials on the plane it is 
assumed that open level lines of such potentials 
have a rather complex geometry and arise only at one 
energy level $\, V ({\bf x }) = E_{0} \, $. Based on this, 
we can see that it is natural to consider a quasiperiodic 
potential as an approximation for a random potential 
if the corresponding interval $\, [c_{1} , c_{2}] \, $ 
for it contracts to a point $\, c_{1} = c_{2} = c_{0} \, $.
 
 It can also be noted that in the case of three quasiperiods 
the situation $\, c_{2} > c_{1} \, $ also entails the presence 
of regular topological (and geometric) properties of open 
potential level lines (in the case $\, N \geq 4 \, $, however, 
a similar connection has not yet been established), which also 
distinguishes such potentials from random ones. On the other hand, 
in the situation $\, c_{1} = c_{2} = c_{0} \, $ potentials with 
three quasiperiods can have very complex open level lines, 
which brings them very close to random ones (see, for example
\cite{DynnBuDA,dynn2,zorich2,Zorich1996,ZorichAMS1997,zorich3,DeLeo1,
DeLeo2,DeLeo3,ZorichLesHouches,DeLeoDynnikov1,dynn4,DeLeoDynnikov2,
Skripchenko1,Skripchenko2,DynnSkrip1,DynnSkrip2,AvilaHubSkrip1,
AvilaHubSkrip2,TrMian,DynHubSkrip}). 
Note, however, that situations (B), (C) and (D) in the case of 
three quasi-periods are impossible and any potential with three 
quasi-periods has at least one unbounded connected level component
at some energy level. 

 In the case $\, N \geq 4 \, $, a situation is possible when 
the potential does not have open level lines at any energy 
level (situation (D)), and closed level lines of arbitrarily 
large sizes arise only at a single level. This property, however, 
is unstable with respect to parallel shifts of the plane $\, \Pi \, $ 
in the space $\, \mathbb{R}^{N} \, $. From the point of view of 
the experimental methods for creating quasiperiodic potentials 
mentioned above, this means that the presence or absence of open 
level lines $\, F |_{\Pi} = c_{0}\, $ can depend on the positions of 
the maxima and minima of periodic potentials superimposed on each other. 
These properties, in our opinion, are not an obstacle to considering 
such potentials as a model of random potentials on the plane.

 A similar situation (the absence of open level lines, as well as of
unbounded connected components of the region of higher or lower 
energies) can be observed in the case of $\, N \geq 4 \, $ also on the 
boundaries of the interval $\, [c_{1} , c_{2 }] \, $ for 
$\, c_{2} > c_{1} \, $ (situations (B) and (C)). As in the 
previous case, this situation is unstable with respect to parallel 
shifts of the plane $\, \Pi \, $ in the space $\, \mathbb{R}^{N} \, $ 
and depends on the positions of the maxima and minima of periodic 
potentials in the experimental techniques we mentioned. It can be 
noted that situations of this type are also of interest from the 
point of view of the theory of percolation in quasiperiodic potentials.

\vspace{1mm}

  Returning to the general consideration, we can note here that the 
intervals $\, [c_{1} , c_{2}] \, $ (or level $\, c_{0}$) represent 
transition regimes between two, in a sense, opposite situations in 
the planes $\, \Pi\, $. Namely, for $\, c < c_{1} (\xi) \, $ each 
plane $\, \Pi \, $ of the direction $\, \xi \, $ can be considered 
as a unique unbounded component of the set $\, F |_{\Pi} > c \, $, 
in which there are ``inclusions'' of regions $\, F |_{\Pi} < c \, $, 
the sizes of which do not exceed some fixed constant 
(for a given $\, c$). (Inside such inclusions there may be smaller 
inclusions of regions $\, F |_{\Pi} > c\, $, etc.). Similarly, 
for $\, c > c_{2} (\xi) \, $, each plane $\, \Pi \, $ of 
the direction $\, \xi \, $ represents an unbounded connected 
component of the set $\, F | _{\Pi} < c \, $ with similar inclusions 
of regions $\, F |_{\Pi} > c \, $, etc. For certain two-dimensional 
systems, in particular, these two situations correspond to different 
signs of the Hall conductivity in the presence of a magnetic field 
orthogonal to the plane of the system.

 The transition between the two situations described above in the 
case of quasiperiodic functions can be quite nontrivial and, 
in particular, may or may not be accompanied by the arising of open 
level lines. As we can see, the last property may be especially 
important for potentials that are most suitable for the role of 
random ones from our point of view ($c_{1} (\xi) = c_{2} (\xi) = c_{0} (\xi)$). 
A complex geometry of the resulting level lines can play a very 
important role both in the semiclassical consideration of transport 
phenomena in such systems, and, in fact, also in a more general 
situation (see, for example, 
\cite{JMathPhys,TitovKatsnelson,QuasiperGas,TwoDimSuperPos}). 

 The values $\, c_{1} $, $\, c_{2} $, as well as $\, c_{0} \, $ 
can be called ``critical'' values of the corresponding potentials, 
since at these values qualitative changes in the picture of the 
potential level lines occur. As can be seen from the results 
formulated above, it is at these values that the behavior of 
the potential level lines can be most interesting.
 
\vspace{1mm}

 Theorem 2.1 directly implies

\vspace{1mm}

\noindent
{\bf Theorem 2.2}

 Let $\, \xi \, $ be a non-special direction in 
$\, \mathbb{R}^{N} \, $ such that 
$\, c_{1} (\xi) < c_{2} (\xi ) \, $. Then for 
any $\, c \in ( c_{1} (\xi) , c_{2} (\xi) ) \, $ each 
plane $\, \Pi \, $ of the direction $\, \xi \, $, which 
is non-singular for the value $\, c \, $, contains open level 
lines $\, F |_{\Pi} = c \, $.

\vspace{1mm}

 Comment.

 In fact, it is not difficult to see that all the statements 
formulated above are also true for planes containing singular 
levels $\, F |_{\Pi} = c \, $ for non-special directions 
$\, \xi \, $ in $\, \mathbb{R}^{N} \, $. In this case, along 
with open level lines, we must consider all non-compact connected 
level components $\, F |_{\Pi} = c \, $ in the plane $\, \Pi \, $, 
and, along with closed ones, compact connected components. 
It can be seen that the above statements, as well as their proofs, 
do not, in fact, undergo essential changes. We just recall that 
we mean here the fulfillment of the general position conditions; 
in particular, we assume that each singular connected component 
$\, F |_{\Pi} = c \, $ for non-special directions $\, \xi \, $ 
contains at most a finite number of multiple saddles of the 
function $\, F |_{\Pi} \, $.

\vspace{1mm}

 It can also be noted that any completely irrational 
direction $\, \xi \, $ is non-special either 
in $\, \mathbb{R}^{N} \, $, or in a space of lower dimension. 
In the latter case, it is natural to consider the corresponding 
functions $\, F |_{\Pi} \, $ as functions with a smaller number 
of quasi-periods.

\vspace{1mm}

 The statements formulated above refer to the most general 
case and describe the situation for completely irrational 
directions $\, \xi $. As we have already said, the 
case $\, N = 3 \, $ admits a more detailed description of 
the level lines of the functions $\, F |_{\Pi} = c \, $. 
Moreover, a more detailed description also exists in the 
case of $\, N \geq 4 \, $ for the partially irrational 
directions $\, \xi \, $ that we defined at the beginning 
of the section (see \cite{DynMalNovUMN}).

\section{Multidimensional generalizations of the Novikov problem}
\setcounter{equation}{0}

 A natural generalization of the Novikov problem is the problem of 
describing level surfaces of quasiperiodic functions in the 
space $\, \mathbb{R}^{n} \, $ with $\, N \, $ quasiperiods. 
Here we naturally call a quasiperiodic function in 
$\, \mathbb{R}^{n} \, $ with $\, N \, $ quasiperiods 
the restriction of an $N$-periodic function 
$\, F (z^{1} , \dots , z^{N}) \, $ onto the space 
$\, \mathbb{R}^{n} \, $ for some embedding 
$\, \mathbb{R}^{n} \subset \mathbb{ R}^{N} \, $ ($n < N$). 
As before, we assume here that the function 
$\, F (z^{1}, \dots , z^{N}) \, $ is periodic with respect to 
the standard integer lattice 
$\, \mathbb{Z}^{ N} \subset \mathbb{R}^{N} \, $. As a general 
position condition, we can, for example, require that the 
subspace $\, \mathbb{R}^{n} \, $ intersects transversally with 
any integer subspace $\, \mathbb{R}^{m} \subset \mathbb{R}^{N} \, $. 
The essential parameters for us will be the direction of 
embedding $\, \mathbb{R}^{n} \subset \mathbb{R}^{N} \, $ 
(i.e. a point of the Grassmann manifold $\, G_{N, n} $), 
as well as the parameters of parallel shifts of 
$\, \mathbb{R}^{n} \, $ in $\, \mathbb{R}^{N} \, $ in directions 
orthogonal to it (i.e. a point in the space $\, \mathbb{R}^{N-n} $).
 
 Certainly, physical applications of the generalized Novikov problem 
relate primarily to the case $\, n = 3 \, $. As in the two-dimensional 
case, here we can especially highlight the case $\, N = 2 n\, $, 
which is most often encountered in the quasicrystals. At the same time, 
the construction of quasiperiodic modulations and potentials of the most 
general form in three-dimensional systems is also an important direction 
in modern experimental physics (see, for example, \cite{Guidoni1}). 
It can be noted that here, as in the two-dimensional case, the study 
of the Novikov problem seems to be the most important in describing 
transport phenomena in such systems.

 As was noted in \cite{DynMalNovUMN} (see arXiv:2306.11257), a number 
of results of this work can, in fact, be generalized to the case of 
quasiperiodic functions in a space of arbitrary dimension. Here we 
will consider in more detail these generalizations, as well as similar 
generalizations of the results presented in the previous section. 

 As in the case of $\, n = 2 \, $, we will be interested here, 
first of all, in the arising of unbounded connected components of 
level surfaces of quasiperiodic functions, as well as their bounded 
components of arbitrarily large size. Here we will call a connected 
component of a level surface (regular or singular) unbounded if 
it cannot be contained in any ball of finite radius in the 
space $\, \mathbb{R}^{n} \, $.

 Below we will assume that level surfaces of quasiperiodic 
functions can be both regular and singular. At the same time, 
we will assume that the corresponding functions 
$\, F |_{\mathbb{R}^{n}} \, $ can only have singularities 
of a certain type, namely, multiple saddles or isolated maxima 
and minima. In addition, here we will assume that for generic 
embedding directions $\, \mathbb{R}^{n} \subset \mathbb{R}^{N} \, $ 
each level of the function $\, F |_{ \mathbb{R}^{n}} \, $ 
can contain only a finite number of its singular points 
(in particular, the number of singular connected components of 
the set $\, F |_{\mathbb{R}^{n}} = c \ , $ is finite for any 
value of $\, c$).

 As in the case of $\, n = 2 \, $, we will consider here the 
generalized Novikov problem for all subspaces 
$\, \mathbb{R}^{n} \subset \mathbb{R}^{N} \, $ of a fixed 
direction $\, \xi \in G_{N,n} \, $. Similar to the case $\, n = 2 \, $ 
(\cite{DynMalNovUMN}), we can formulate here the following lemma.

\vspace{1mm}

\noindent
{\bf Lemma 3.1}

 Let $\, F (z^{1}, \dots, z^{N}) \, $ be an $N$-periodic function  
in $\, \mathbb{R}^{N} \, $ and let $\, \xi \in G_{N,n} \, $ 
and $\, c \in \mathbb{R} \, $ be such that for any $n$-dimensional 
affine subspace $\, V \, $ of direction $\, \xi \, $ all level 
surfaces $\, F |_{V} = c \, $ (singlar or nonsingular)
are compact. Then the diameters of all these level surfaces are bounded 
from above by one constant common to all the subspaces of direction 
$\, \xi \, $.

\vspace{1mm}

 Proof.

 Like in the case $\, n = 2 \, $, we can claim that each compact level 
surface $\, F |_{V} = c \, $ (both singular and non-singular) 
has a neighborhood of finite diameter such that all other level surfaces 
$\, F |_{V} = c \, $ in parallel subspaces that intersect this neighborhood 
lie entirely in it. Since the hypersurface 
$\, F (z^{1}, \dots, z^{N}) = c \, $ is compact in the torus 
$\, \mathbb{T}^{N} \, $, we can choose a finite family of such neighborhoods
so that their images cover this hypersurface in $\, \mathbb{T}^{N} \, $. 

\hfill{Lemma 3.1 is proved.}

\vspace{1mm}

\noindent
{\bf Lemma 3.2}

 Let $\, F (z^{1}, \dots, z^{N}) \, $ be an $N$-periodic function, 
and let $\, \xi \in G_{N,n} \, $ be a fixed $n$-dimensional direction 
in $\, \mathbb{R}^{N} \, $. Then the set of values $\, c \in \mathbb{R} \, $ 
such that for at least one subspace $\, V \, $ of direction $\, \xi \, $ the 
level surface $F |_{V} = c \, $ has unbounded connected components forms 
a closed interval $\, [c_1,c_2] \, $ or consists of a single 
point $\, c_{0} \, $. 

\vspace{1mm}

 The proof of Lemma 3.2 follows the pattern of the proof of Theorem 1 in 
\cite{dynn3}, which only needs to be slightly modified using Lemma 3.1
and applied to the multidimensional case. 

\vspace{1mm}

 By analogy with the case $\, n = 2 \, $, let us also formulate here 
the following lemma. 

\vspace{1mm}

\noindent
{\bf Lemma 3.3}

 For any generic direction $\, \xi \in G_{N,n} \, $ and any 
$\, c \in [c_{1} (\xi) , c_{2} (\xi)] \, $ each affine subspace $\, V \, $ 
of direction $\, \xi \, $ contains either unbounded connected 
components (singular or non-singular) of the set $\, F |_{V} = c \, $, 
or bounded non-singular connected components of this set of arbitrarily 
large size (or both). 

\vspace{1mm}

 Proof.

 The proof of Lemma 3.3 is similar to the proof of the analogous 
statement for the case $\, n = 2 \, $ (\cite{DynMalNovUMN}, arXiv:2306.11257). 
Let us have an affine subspace $\, V \, $ of direction $\, \xi \, $, in
which there are no unbounded connected components of the 
set $\, F |_{V} = c \, $, and the sizes of all its regular connected 
components are limited by one constant. By virtue of the definition of 
a generic direction, the sizes of all connected components of the set 
$\, F |_{V} = c\, $ in this case, in fact, are also limited by one constant.
(By the size of a component we mean the minimum diameter $\, D \, $ of the 
sphere $\, S^{n-1} \, $ in which it can be placed). 

 Let us take a subspace $\, V^{\prime} \, $ of direction $\, \xi \, $, 
containing an unbounded component (singular or non-singular) of the surface
$\, F|_{V^{\prime}} = c \, $. According to our general assumptions,
we can take a connected part of this component such that it does not 
contain singular points and cannot be enclosed in a sphere of diameter $D$. 
There is a small $\delta$ such that this part is preserved for all parallel 
shifts of $\, V^{\prime} \, $ in all transversal directions by a distance 
less than $\delta$.

 Consider the corresponding $\delta$-neighborhood $O$ of $\, V^{\prime} \, $. 
Integer shifts of the original subspace $\, V \, $ are everywhere 
dense in $\, \mathbb{R}^{N} \, $ for our $\, \xi \, $ and must fall in 
the neighborhood $O$. We get a contradiction.

\hfill{Lemma 3.3 is proved.}

\vspace{1mm}

 We will now present analogues of the statements made in the previous 
section for the multidimensional generalization of the Novikov problem. 
We will not now assume that the subspaces $\, V \, $ are non-singular, 
since the proofs of our assertions do not change significantly for subspaces 
containing singular points of the functions $\, F |_{V} \, $
under our general assumptions.
 
 As in the case $\, n = 2 \, $,  for each affine subspace $\, V \, $ 
and a fixed value $\, c \, $ denote by $\, \Omega^{-}_{c} (V) \, $ 
the set of points in $\, V \, $, where $\, F |_{V} < c \, $. Similarly, 
denote by $\, \Omega^{+}_{c} (V) \, $ the set of points in $\, V \, $, 
where $\, F |_{V} > c \, $. Under our general assumptions, we can again 
formulate here the following simple lemma.

\vspace{1mm}

\noindent
{\bf Lemma 3.4}

 Let $\, \xi \, $ be a generic $n$-dimensional direction in 
$\, \mathbb{R}^{N} \, $ and $\, V \, $ be an affine subspace
of direction $\, \xi \, $. Then the presence of unbounded 
connected components of the level surface $\, F |_{V} = c \, $ 
(singular or non-singular) is equivalent to the existence of unbounded 
connected components of both sets $\, \Omega^{-}_{c} (V) \, $ 
and $ \, \Omega^{+}_{c} (V) \, $ in $\, V \, $. 

\vspace{1mm}
 
 By analogy with the two-dimensional case let us call here a 
non-singular bounded component of the surface $\, F |_{V} = c \, $ 
an ``electronic'' component if the region $\, \Omega^{+}_{c} (V) \, $ 
adjoins it from the outside. Similarly, we call a non-singular bounded 
component of the surface $\, F |_{V} = c \, $ a ``hole'' component 
if the region $\, \Omega^{-}_{c} (V)  \, $ adjoins it from the outside.

 Similar to the two-dimensional case, we can formulate the following 
lemmas here.

\vspace{1mm}

\noindent
{\bf Lemma 3.5}

 Let $\, \xi \, $ be a generic $n$-dimensional direction in 
$\, \mathbb{R}^{N} \, $ and $\, V \, $ be an affine subspace
of direction $\, \xi \, $. Let 
$\, c \in [c_{1} (\xi) , c_{2} (\xi) ] \, $
and the size of the electronic components of $\, F |_{V} = c \, $ 
in $\, V \, $ is bounded above by one constant. Then 
in $\, V \, $ there are unbounded connected components of 
the set $\, \Omega^{-}_{c} (V) \, $.

\vspace{1mm}

\noindent
{\bf Lemma 3.5$^{\prime}$}

 Let $\, \xi \, $ be a generic $n$-dimensional direction in 
$\, \mathbb{R}^{N} \, $ and $\, V \, $ be an affine subspace
of direction $\, \xi \, $. Let 
$\, c \in [c_{1} (\xi) , c_{2} (\xi) ] \, $
and the size of the hole components of $\, F |_{V} = c \, $ 
in $\, V \, $ is bounded above by one constant. Then 
in $\, V \, $ there are unbounded connected components of 
the set $\, \Omega^{+}_{c} (V) \, $. 

\vspace{1mm}

 The proof of Lemmas 3.5 and 3.5$^{\prime}$ is completely similar 
to the case $\, n = 2 \, $ taking into account Lemma 3.3. Using 
the lemmas formulated above, we can state here that in a subspace 
$\, V \, $ of a generic direction $\, \xi \, $ and 
$\, c \in [c_{1 } (\xi) , c_{2} (\xi) ] \, $, under our general 
assumptions, only one of the following four situations is possible.

\vspace{1mm}

\noindent
(A$^{\prime}$) The subspace $\, V \, $ contains unbounded connected 
components of the surface $\, F |_{V} = c \, $ 
(singular or non-singular), as well as unbounded connected components 
of both sets $\, \Omega^{-}_ {c} (V) \, $ 
and $\, \Omega^{+}_{c} (V) \, $.

\vspace{1mm}

\noindent
(B$^{\prime}$) In the subspace $\, V \, $, only the set 
$\, \Omega^{+}_{c} (V) \, $ has unbounded connected components 
and there are no unbounded connected components of 
the surface $\, F |_ {V} = c \, $. In 
this case $\, V \, $ must contain arbitrarily large non-singular
bounded connected components of $\, F |_{V} = c \, $ of the 
electronic type.

\vspace{1mm}

\noindent
(C$^{\prime}$) In the subspace $\, V \, $, only the set 
$\, \Omega^{-}_{c} (V) \, $ has unbounded connected components 
and there are no unbounded connected components of 
the surface $\, F |_ {V} = c \, $. In 
this case $\, V \, $ must contain arbitrarily large non-singular
bounded connected components of $\, F |_{V} = c \, $ of the 
hole type.

\vspace{1mm}

\noindent
(D$^{\prime}$) The subspace $\, V \, $ contains neither unbounded 
connected components of the set $\, \Omega^{-}_{c} (V) \, $, nor 
unbounded connected components of the set 
$\, \Omega^{+}_ {c} (V) \, $, nor unbounded connected components 
of the level surface $\, F |_{V} = c \, $. In this case $\, V \, $ 
must contain arbitrarily large non-singular bounded connected 
components of $\, F |_{V} = c \, $ of both electronic and hole types.

\vspace{1mm}

 Here we can also note that for every regular bounded connected 
component $\, F |_{V} = c \, $ there always exists a (small) $\delta$ 
such this component is preserved for all parallel shifts 
$\, \sigma_{\bf a} [V] \, $ in all transversal directions in 
$\, \mathbb{R}^{N} \, $ to distances less than $\delta$. From here 
it is not difficult to see, in fact, that for generic directions 
$\, \xi \in G_{N,n} \, $, the existence of arbitrarily large regular 
bounded components of the set $\, F |_{V} = c \, $ of a given type 
implies their existence also in all other subspaces $\, V^{\prime} \, $ 
of this direction. It follows, in particular, that if for some values 
of $\, c \, $ ($c = c_{1} (\xi)$, $c = c_{2} (\xi)$ or $c = c_{ 0} (\xi)$) 
unbounded connected components of the set $\, F |_{V} = c \, $ arise only in 
some (not all) subspaces $\, V \, $ of a generic direction $\, \xi \, $, 
they necessarily arise together with arbitrarily large bounded regular 
components of this set.

\vspace{1mm}

 It is also easy to see that in case (D$^{\prime}$), due to the 
absence of unbounded connected components of the sets 
$\, \Omega^{-}_{c} (V) \, $ and $\, \Omega^{+}_{ c} (V) \, $, 
each electronic component $\, F |_{V} = c \, $ must be contained 
in some larger hole component. In turn, each hole component 
$\, F |_{V} = c \, $ must also be contained in some 
larger electronic component $\, F |_{V} = c \, $.

\vspace{1mm}

 Finally, as in the case $\, n = 2 \, $, we can formulate here the 
following theorems.

\vspace{1mm}

\noindent
{\bf Theorem 3.1}

 Let $\, \xi \, $ be a generic $n$-dimensional direction 
in $\, \mathbb{R}^{N} \, $, 
$\,\, c \in [c_{1} (\xi) , c_{ 2} (\xi) ] \, $, 
and $\, V \, $ be a subspace of direction $\, \xi \, $.  
Then in the subspace $\, V \, $:

\vspace{1mm}

\noindent
1) Situation (B$^{\prime}$) is possible only for $\, c = c_{1} (\xi) \, $.

\vspace{1mm}

\noindent
2) Situation (C$^{\prime}$) is possible only for $\, c = c_{2} (\xi) \, $.

\vspace{1mm}

\noindent
3) Situation (D$^{\prime}$) is possible only for 
$\, c = c_{0} (\xi) = c_{1} (\xi) = c_{2}(\xi) \, $.

\vspace{1mm}

\noindent
{\bf Theorem 3.2}

 Let $\, \xi \in G_{N,n} \, $ be a generic direction in 
$\, \mathbb{R}^{N} \, $ such that 
$\, c_{1} (\xi) < c_{2} (\xi ) \, $. Then for 
any $\, c \in ( c_{1} (\xi) , c_{2} (\xi) ) \, $ each 
subspace $\, V \, $ of direction $\, \xi \, $ contains  
unbounded connected components (singular or non-singular)
of the suraface $\, F |_{V} = c \, $.

\vspace{1mm}

 The proof of Theorems 3.1 and 3.2 is carried out similarly 
to the case $\, n = 2 \, $, taking into account the above lemmas.

\section{Conclusion}
\setcounter{equation}{0}

  The paper considers questions related to the general Novikov problem, 
i.e. the description of the level lines of quasi-periodic functions on the 
plane with an arbitrary number of quasi-periods. More precisely, the 
cases of arising of open level lines of such functions, as well as their 
closed level lines, having arbitrarily large sizes, are investigated.  
The arising of such level lines is closely related to certain values 
of the function under consideration, lying within the full range 
of its values. The latter circumstance actually plays an important 
role in the study of dynamical systems related to the Novikov problem, 
where the considered function $\, F\, $ often plays the role 
of the Hamiltonian function. In fact, the presented results can also 
be formulated for the generalized Novikov problem, namely, the problem 
of describing level surfaces of quasiperiodic functions in $n$-dimensional 
spaces. As a rule, the geometric features of such level surfaces appear 
primarily in describing transport phenomena in systems with 
quasi-periodic modulations.

\end{document}